\documentclass[aps,prd,preprint,showkeys,showpacs]{revtex4-1}
\usepackage{graphicx}                                
\usepackage{amsmath,amsfonts,bbm}
\graphicspath{{./Figures/}}                          
\usepackage[caption=false]{subfig}
\usepackage{epsfig}
\usepackage{amssymb}
\usepackage{amsfonts}
\usepackage{mathrsfs}
\newcommand{\Tr}{\mbox{Tr}}

\newcommand{\beq}{\begin{equation}}
\newcommand{\eeq}{\end{equation}}
\newcommand{\bea}{\begin{eqnarray}}
\newcommand{\eea}{\end{eqnarray}}
\newcommand{\beas}{\begin{eqnarray*}}
\newcommand{\eeas}{\end{eqnarray*}}
\newcommand{\eq}{\begin{equation}}
\newcommand{\en}{\end{equation}}
\newcommand{\eqa}{\begin{eqnarray}}
\newcommand{\ena}{\end{eqnarray}}

\newcommand{\vek}[1]{\mathbf{#1}}

\begin{document}

\title{Ghost propagator and the Coulomb form factor from the lattice}
\author{G.~Burgio}
\author{M.~Quandt}
\author{H.~Reinhardt}
\affiliation{Institut f\"ur Theoretische Physik, Auf der Morgenstelle 14, 
72076 T\"ubingen, Germany}

\begin{abstract}
We calculate the Coulomb ghost propagator $G(|\vek{p}|)$ and the static 
Coulomb potential $V_C(|\vek{r}|)$ for $SU(2)$ Yang-Mills theory on the 
lattice. In view of possible scaling violations related to deviations from the 
Hamiltonian limit we use anisotropic lattices to improve the temporal 
resolution. We find that the ghost propagator is infrared enhanced with an 
exponent $\kappa_{\rm gh} \gtrsim 0.5$ while the Coulomb potential exhibits a 
string tension larger than the Wilson string tension, 
$\sigma_C \simeq 2 \sigma$. This agrees with the Coulomb ``scaling'' scenario 
derived from the Gribov-Zwanziger confinement mechanism.

\end{abstract}

\maketitle

\section{Introduction}

Yang-Mills theories in Coulomb gauge have attracted increasing interest in the 
last years, both in the continuum and on the lattice. Non-perturbative analytic 
predictions based on the Gribov-Zwanziger confinement mechanism 
\cite{Gribov:1977wm,Zwanziger:1995cv} turn out to be concise and elegant in 
this gauge and numerical simulations have confirmed them so far. 

In covariant gauges the Gribov-Zwanziger approach requires a restriction of 
the functional integral beyond the Faddeev-Popov mechanism 
\cite{Faddeev:1967fc}, which has long been known to be insufficient to extract 
a 
unique field representative along the gauge orbit and thus cannot be used to 
define the partition function beyond perturbation theory 
\cite{Gribov:1977wm,Singer:1978dk,Killingback:1984en}. As a remedy one adds 
extra terms to the action which are, in general, non-local and often referred 
to as the \emph{Horizon function} or \emph{Horizon condition}. The purpose of 
these extra terms is to limit the fields in the functional integral
to the so-called fundamental modular region $\Lambda$, which contains only 
\emph{absolute} minima of the gauge fixing functional and thus 
eliminates 
the over-counting of gauge copies from the same orbit (except for topologically 
non-trivial copies).

In practice, however, it is almost impossible to limit the functional integral 
beyond the so-called first Gribov region $\Omega \supseteq \Lambda$, where the 
Faddeev-Popov operator is positive definite. Moreover, any restriction on the 
integration range imposed other than through the Faddeev-Popov mechanism will 
break the Becchi-Rouet-Stora-Tyutin (BRST) symmetry 
\cite{Becchi:1975nq,Tyutin:1975qk}. 
Whether such enlarged action including Horizon terms exhibits some other 
symmetry and what the possible consequences in covariant gauges are 
(e.g.~in terms of contributions from non-standard condensates 
\cite{Burgio:1997hc,Akhoury:1997by,Boucaud:2000ey,Dudal:2010tf}) 
is still under active debate, 
cfr.~Ref.~\cite{Vandersickel:2012tz,Dudal:2012sb} and 
references therein for recent results.

In contrast to the situation in covariant gauges sketched above, the physical 
implications 
of the Gribov-Zwanziger approach are quite transparent in \emph{Coulomb gauge}.
In the Hamiltonian formulation 
\cite{Christ:1980ku,Szczepaniak:2001rg,Feuchter:2004gb,Feuchter:2004mk},
for instance, the Gribov-Zwanziger idea amounts to a mere projection of the 
physical Hilbert space onto the subspace of states satisfying Gau\ss's 
law \cite{Reinhardt:2008pr,Reinhardt:2008ij}.  
Besides being conceptually simpler, this has at least two advantages:
\begin{itemize}
 \item Unlike the Kugo-Ojima approach \cite{Kugo:1979gm} in covariant gauges,
where the existence of a globally conserved BRST charge 
$Q_{\rm BRST} |\Psi\rangle_{\mbox{\scriptsize ph}} = 0$
is essential, the Hamiltonian approach in Coulomb gauge requires no additional 
assumptions to ensure Gau\ss's law, i.e. a vanishing colour charge on physical 
states, $Q_{c} |\Psi\rangle_{\mbox{\scriptsize ph}} = 0$. 
\item The construction of the physical Hilbert space in Coulomb gauge is much 
simpler than in the fully gauge-invariant Hamiltonian 
approach \cite{Burgio:1999tg}. In fact, the consequences of Gau\ss's law in 
Coulomb gauge 
can be implemented \emph{exactly} in a functional integral, which is therefore 
well suited for 
further approximations without the impediment of additional constraints such 
as the conservation of $Q_{\rm BRST}$. 
\end{itemize}
The Horizon condition in Coulomb gauge implies both a \emph{static}
(i.e.~equal-time) transverse gluon propagator \cite{Gribov:1977wm}
which vanishes at zero momentum, and a ghost dressing function which diverges 
in the same limit  \cite{Zwanziger:1995cv}. Physically, the last statement 
means 
that the Yang-Mills vacuum behaves as a perfect color dia-electric medium 
(i.e.~a dual superconductor), because the dielectric function of the Yang-Mills 
vacuum agrees with the inverse ghost dressing function  \cite{Reinhardt:2008ek}.
As a consequence the dual Meissner effect, which has long been a model for the 
origin of the confining force in so-called Abelian gauges, can also be 
applied to Coulomb gauge. A natural quantity
to study confinement in Coulomb gauge is the non-Abelian Coulomb potential,
which provides an upper bound for the quark-antiquark free energy 
$V(r) \leq V_C(r)$, i.e.~there 
is \emph{no confinement without Coulomb confinement} \cite{Zwanziger:2002sh}.

For all these reasons Coulomb gauge is best suited for direct investigations 
of the QCD wave functional. After pioneering analytical 
\cite{Drell:1978hr,Greensite:1979yn,Arisue:1982tt} and numerical work 
\cite{Velikson:1984qw}, 
such studies have experienced a broad popularity in the literature, 
cfr.~\cite{Hecht:1992ps,Kogan:1994wf,Karabali:1995ps,Brown:1997nz,Brown:1997gm,Karabali:1998yq,Leigh:2005dg,Greensite:2007ij,Quandt:2010yq,Reinhardt:2011fq}.
In particular, the Hamiltonian approach lends itself to variational 
formulations \cite{Szczepaniak:2001rg,Feuchter:2004mk,Velikson:1984qw}, which 
allow to address non-perturbative problems in Yang-Mills theory in a 
rather direct and concise way. A main ingredient in these techniques are 
static (equal-time) two-point functions, so that a direct investigation 
of the Gribov-Zwanziger scenario for such correlators is important.

There have been a number of studies of this subject in the 
literature \cite{Langfeld:2004qs,Voigt:2007wd,Nakagawa:2008zzc,Nakagawa:2009zf}.
From previous analyses of the gluon propagator in Coulomb 
gauge \cite{Burgio:2008jr,Burgio:2008yg,Burgio:2009xp,Burgio:2012ph} it is 
known 
that the difficulties in reaching the Hamiltonian limit on a lattice with 
a finite time-resolution can have drastic consequences: the resulting 
scaling violations, if untreated, prevent a multiplicative renormalization 
of the gluon propagator and also affect the correct analysis in the 
deep infrared. We therefore compute the Green 
functions on \emph{anisotropic} lattices with a high temporal resolution and 
study the 
effects of approaching the Hamiltonian limit. We also employ a high quality 
of gauge fixing, particularly important for the ghost propagator. 

The paper is organized as follows: in the next section, we discuss the exact 
definition of the relevant correlators, both in the continuum and on the 
lattice.
We will also display the expectations on the conformal behaviour of the 
correlators
in the deep-infrared, as raised by variational and functional methods.
Section 3 contains a description of our numerical setup, as well as a detailed
discussion of our results for both the two-point functions and the Coulomb 
potential.
We also compare to continuum methods and give arguments to explain possible 
deviations. Finally, we conclude in section 4 with a brief summary and outlook.

\section{Correlators in Coulomb gauge}
\subsection{Correlators in the continuum}

At fixed time $t$, the Coulomb gauge condition $\vek{\nabla} \cdot \vek{A} = 0$
is complete (up to possible singularities and Gribov ambiguities,
see Ref.~\cite{Jackiw:1977ng}), i.e.~any residual gauge symmetry is 
space-independent and thus acts as a global gauge at fixed time $t$. 
In this situation, we are interested in the static (i.e.~equal time) 
propagators,
\begin{eqnarray}
D(\vek{p}) &=& \frac{\delta^{ab}\,\delta_{ij}}{2\,N_A}\,
    \big\langle {A}^a_{i}(\vek{p},t) {A}^b_{j}(-\vek{p},t)
    \big\rangle = \frac{\delta^{ab}\,\delta_{ij}}{2 \,N_A}\,
    \int \frac{d\,p_0}{2\pi}\,\big\langle {A}^a_{i}(p) 
    {A}^b_{j}(-p)\big\rangle\label{stgl}\\[2mm]
G(\vek{p}) &=& \frac{d(\vek{p})}{|\vek{p}|^{2}} = \frac{\delta^{ab}}{N_A}\,
    \big\langle {\bar{c}}^a(\vek{p}) {c}^b(-\vek{p})
    \big\rangle =  \mathrm{tr}\,\big\langle\,  (- \vek{D} \cdot \vek{\nabla})^{-1}\,\big\rangle
     \label{stgh}\\[2mm]
D_0(\vek{p}) &=& \frac{\delta^{ab}}{N_A} \,\big\langle {A}^a_{0}(\vek{p},t) 
{A}^b_{0}(-\vek{p},t) \big\rangle\,\sim\label{tprop}\\[2mm]
 &\sim& V_C(\vek{p}) \equiv g^2 \,\mathrm{tr}\,\big\langle  
    (- \vek{D} \cdot \vek{\nabla})^{-1}
    (-\vek{\nabla}^2) (- \vek{D} \cdot \vek{\nabla})^{-1}\big\rangle\,,
\label{stpot}
\end{eqnarray}
where in the last equation contributions from non contact terms have been 
dropped \cite{Zwanziger:1998ez}. $N_A$ is the dimension of the adjoint colour 
representation and 
$\mathrm{tr}$ is the normalized colour trace; we use the colour group 
$G=SU(2)$ with 
$N_A = 3$ throughout this paper. Moreover, the covariant derivative in the adjoint 
representation reads $\vek{D}^{ab} \equiv \nabla\,\delta^{ab} + g \epsilon^{abc} \vek{A}^c$, 
and $(- \vek{D}\cdot \nabla)$ is the Faddeev-Popov operator in Coulomb gauge.  

Within the  Hamiltonian variational approach one finds 
two sets of solutions for the gap equations, 
called ``critical'' \cite{Feuchter:2004mk,Epple:2006hv} 
and ``subcritical'' \cite{Epple:2007ut}. Here we will be mainly interested in 
the 
critical solution, which is also believed to be the physically relevant one.
In this case one finds a Coulomb potential Eq.~(\ref{stpot}) which rises 
linearly as a function of the distance \cite{Epple:2006hv}.
Furthermore, the gluon and ghost propagators in Eq.~(\ref{stgl},\ref{stgh}) 
exhibit a 
conformal scaling in the deep infrared, i.e.~they behave as a power of momentum
$|\vek{p}| \ll 1$, 
\begin{equation}
D(\vek{p}) \sim \frac{1}{|\vek{p}|^{\kappa_{\rm gl}}} \,,\qquad\qquad
d(\vek{p}) \sim \frac{1}{|\vek{p}|^{\kappa_{\rm gh}}}\,.
\label{IRexp}
\end{equation} 
Similarly, asymptotic freedom indicates that the large momentum behaviour of
the 
propagators essentially follows from the bare action, modified by logarithmic 
corrections with appropriate anomalous dimensions,
\begin{eqnarray}
D(\vek{p}) \sim \frac{1}{|\vek{p}| \log^{\gamma_{\rm gl}}{\frac{|\vek{p}|}{m}}}\,,\qquad\quad
d(\vek{p}) \sim \frac{1}{\log^{\gamma_{\rm gh}}{\frac{|\vek{p}|}{m}}}\,. 
\label{UVexp}
\end{eqnarray} 
In both regimes, the relevant exponents are further constrained by the 
assumption that the (static) ghost-gluon vertex is essentially trivial. 
i.e.~proportional to the bare vertex, for any kinematical configuration. 
This leads directly to the so-called \emph{sum rules} \cite{Epple:2007ut}
\begin{eqnarray}
\kappa_{\rm gl} + 2 \kappa_{\rm gh} &=& 1\nonumber\\
\gamma_{\rm gl} + 2 \gamma_{\rm gh} &=& 1\,.
\label{sumrules}
\end{eqnarray} 
The two solutions of the variational approach are derived under the same 
assumption, with the specific values for the 
exponents \cite{Feuchter:2004mk,Schleifenbaum:2006bq,Epple:2006hv,Epple:2007ut}:
\begin{eqnarray}
\kappa^c_{\rm gl} = -1&\qquad& \kappa^c_{\rm gh}  = 1 \\
\kappa^s_{\rm gl} \simeq -0.6&\qquad& \kappa^s_{\rm gh}  \simeq 0.8 \\
\gamma_{\rm gl} = 0&\qquad&\gamma_{\rm gh} = \frac{1}{2}\,.
\label{NPexp}
\end{eqnarray} 
Notice that Eq.~(\ref{IRexp}) implies, in general, an infrared mass generation 
for both gluon and ghost, unless $\kappa_{\rm gl} =  1$ and $\kappa_{\rm gh} = 0$, 
their tree-level values. Such an infrared behaviour agrees with the original 
analysis of Gribov \cite{Gribov:1977wm} and, in particular, implies an 
infrared vanishing static gluon propagator $D(\vek{p}) \to 0$ and an infrared 
divergent ghost form factor $d(\vek{p}) \to \infty$, as $|\vek{p}| \to 0$. 
Physically, these findings translate into a diverging gluon self-energy 
$\omega(\vek{p}) \sim D(\vek{p}) ^{-1}$, and a vanishing dielectric function 
of the Yang-Mills vacuum, $\epsilon(\vek{p}) \sim d(\vek{p}) ^{-1}$, which in 
turn implies dual superconductivity \cite{Reinhardt:2008ek}.
In Ref.~\cite{Leder:2010ji} the ghost and gluon propagator were also calculated
from the renormalization group flow equations, assuming a bare ghost-gluon
vertex and UV propagators with vanishing anomalous dimensions 
($\gamma_{\rm gl} = \gamma_{\rm gh} = 0$); the corresponding IR exponents 
$\kappa_{\rm gl} \simeq  -0.28$ and $\kappa_{\rm gh} \simeq 0.64$ turn out to be
somewhat smaller than in Eq.~(\ref{NPexp}).

The large momentum behaviour cited above for both propagators shows interesting 
non-perturbative effects as well. Despite asymptotic freedom (and in contrast 
to Landau gauge), the predictions Eq.~(\ref{NPexp}) for the anomalous 
dimensions cannot be directly compared to perturbation theory 
\cite{Watson:2007mz,Watson:2007vc}, since no obvious rainbow-ladder-like 
re-summation technique is available in Coulomb gauge. 
In fact, it is not even clear whether static propagators can be accessed in 
standard perturbation theory in the first place: the Slavnov-Taylor identities 
imply a highly non-trivial energy (i.e.~$p_0$) dependence in the Green's 
functions \cite{Watson:2008fb}, which could spoil the naive static limit and 
thus require a non-perturbative treatment anyhow 
\cite{Watson:2010cn,Andrasi:2010dv,Andrasi:2012wg}. Alternatively, a 
perturbative approach based on the functional 
renormalization group has been attempted in 
Ref.~\cite{Campagnari:2009km,Campagnari:2009wj}; their prediction 
$\gamma_{\rm gl} = 3/11$ and $\gamma_{\rm gh} =4/11$ 
differs, however, quite substantially from the non-perturbative results in 
Eq.~(\ref{NPexp}).

\subsection{Correlators on the Lattice}

A comparison between continuum and lattice Coulomb propagators was pioneered in 
Ref.~\cite{Cucchieri:2000gu}; later studies with different techniques gave, 
however, contradicting results 
\cite{Langfeld:2004qs,Voigt:2007wd,Nakagawa:2008zzc,Nakagawa:2009zf}. 
The difficulty lies in the fact that most continuum results for static 
quantities are naturally obtained in the gauge $A_0 = 0$, which cannot be attained 
directly in lattice calculations because of the compactified time direction. 
Furthermore, any finite lattice has a finite time resolution and the 
associated discretization artifacts preclude
the Hamiltonian limit and give rise to severe renormalization problems.
A way to circumvent these issues was first proposed in  
Ref.~\cite{Burgio:2008jr,Burgio:2008yg,Burgio:2009xp}. There it was shown that 
the static gluon propagator of Eq.~(\ref{stgl}) agrees, after dealing with 
compactification and renormalization artifacts, with the exponents in 
Eq.~(\ref{NPexp}). More precisely, the static gluon propagator in $D=3+1$ was 
shown to be surprisingly well described, over the whole momentum range, by 
Gribov's original proposal \cite{Gribov:1977wm}:
\beq
D(\vek{p}) = \frac{|\vek{p}|}{2\,\sqrt{|\vek{p}|^4+M^4}}\,.
\label{gribov}
\eeq
As argued in Ref.~\cite{Burgio:2008jr,Burgio:2008yg}, the problem in the 
``naive'' extraction of $D(\vek{p})$ from the lattice data lies in the 
standard, isotropic discretization of Yang-Mills theories. The use of 
anisotropic lattices, which approach the Hamiltonian limit 
\cite{Burgio:2003in} for large anisotropy
$\xi = a_s/a_t$ (see Sec.~\ref{setup}), was therefore proposed as a general 
tool in Coulomb gauge investigations. Further independent studies in 
Ref.~\cite{Nakagawa:2009is,Nakagawa:2011ar} have confirmed the presence of 
such effects and the improvement of lattice results for $\xi >1$ for the
gluon correlators. Incidentally, the authors there
were still not able to extract a coulomb string tension from the
analysis of the temporal propagator, Eq.~(\ref{tprop}); the supposed 
equivalence between Eq.~(\ref{tprop}) and Eq.~(\ref{stpot})
\cite{Zwanziger:1995cv,Zwanziger:2002sh} remains therefore an open issue.

In this paper we will continue and complete the lattice analysis 
of Green's functions in Coulomb gauge for pure $SU(2)$ Yang-Mills theory at 
$T=0$,
which was started in Ref.~\cite{Burgio:2008jr,Burgio:2009xp}. 
We will concentrate on the ghost form factor $d(\vek{p})$ and the Coulomb
potential $V_C(\vek{p})$, and study, in particular, the 
anisotropy effects and the validity of the confinement scenario sketched above. 
A study of the Coulomb gauge quark propagator with light dynamical fermions 
has been published elsewhere \cite{Burgio:2012ph}.

\section{Results}
\subsection{Numerical setup}
\label{setup}
Pure Yang-Mills theory, as any continuum field theory, can be formulated
in the Hamiltonian picture \cite{Kogut:1974ag}, where space and time are 
treated separately. Upon discretization, simple RG-group arguments 
\cite{Kogut:1979vg} indicate that for any given isotropic lattice version of 
the field theory in question a corresponding anisotropic counterpart lying 
in the same universality class exists \cite{Svetitsky:1982gs}. The latter 
describes the same physics, albeit with two different lattice spacings $a_s$ 
and $a_t$ for the space and time directions. For Yang-Mills actions in $D=d+1$ 
dimensions built in terms of the character $\chi$ of $m \times n$ Wilson loops 
$P_{\mu\nu}(x;m,n)$, 
\beq
S(\chi;m,n) = \beta^\chi \sum_x \sum_{\mu>\nu=1}^{d+1} \left(1-
\frac{1}{\mbox{dim}(\chi)}
\chi\left[P_{\mu\nu}(x;m,n)\right]\right)
\label{bla1}
\eeq
the anisotropic counterpart will read (see Ref.~\cite{Burgio:2003in} and 
references therein):
\bea
S'(\chi;m,n) &=& \sum_x \Bigg\{\beta_s^\chi \sum_{i> j=1}^d\left(1-
\frac{1}{\mbox{dim}(\chi)}\chi\left[P_{ij}(x;m,n)\right]\right)\nonumber\\[2mm]
&&\quad{}+\beta_t^\chi 
\sum_{i=1}^d
\left(1-\frac{1}{\mbox{dim}(\chi)}\chi\left[P_{i,d+1}(x;m,n)\right]\right)\Bigg\}\,.
\label{bla2}
\eea
For each choice of $\beta_s^\chi \neq \beta_t^\chi$ the two lattice spacings $a_s$ 
and $a_t$ have to be determined non-perturbatively. The couplings are
usually parametrized as $\beta_s^\chi = \beta^\chi\cdot \gamma$ and 
$\beta_t^\chi = \beta^\chi\cdot \gamma^{-1}$, where $\beta^\chi$ is a common 
coupling factor, while $\gamma$ is the \emph{bare} anisotropy; it is related 
to the true (renormalized) anisotropy $\xi = a_s/a_t$ through the 
renormalization constant $\eta$, i.e.~we have $\xi = \gamma \cdot \eta$. The 
non-perturbative value of $\eta$ can be shown to slowly approach, in the 
weak coupling limit, its perturbative expression 
\cite{Burgio:2003in,Burgio:2003nk}.
In much the same fashion, the transition from an isotropic action such as 
eq.~(\ref{bla1}) to its anisotropic counterpart eq.~(\ref{bla2}) can be 
generalized to lattice 
actions of the form $S = \sum_{\chi,mn} c_{\chi,mn}\,S(\chi;m,n)$.
Such anisotropic models are very useful in lattice applications, see 
e.g.~Ref.~\cite{Burgio:2003in,Morningstar:1999rf} and references therein.

Turning to our specific case, we will concentrate on the standard Wilson 
one-plaquette action in 3+1 dimensions, which for $SU(N_c)$ pure gauge theory 
reads:
\beq
S_W = \sum_x \beta \sum_{\mu>\nu=1}^4 \left(1-\frac{1}{N_c}
\mathsf{Re}\left[\Tr\left(P_{\mu\nu}(x)\right)\right]\right)\,,
\eeq
while its anisotropic version is given by:
\bea
S'_W &=& \sum_x \left\{\beta_s \sum_{i>j=1}^3\left(1-
\frac{1}{N_c}\mathsf{Re}\left[\Tr\left(P_{ij}(x)\right)\right]\right)
+\beta_t 
\sum_{i=1}^3\left(1-\frac{1}{N_c}
\mathsf{Re}\left[\Tr\left(P_{i4}(x)\right)\right]\right)\right\}\\
     &=& \beta \sum_x \left\{\gamma \sum_{i>j=1}^3\left(1-
\frac{1}{N_c}\mathsf{Re}\left[\Tr\left(P_{ij}(x)\right)\right]\right)
+\frac{1}{\gamma}
\sum_{i=1}^3\left(1-\frac{1}{N_c}
\mathsf{Re}\left[\Tr\left(P_{i4}(x)\right)\right]\right)\right\}\,.\nonumber
\eea
The values for $a_t$ and $a_s = a_t \xi$ for various choices of $\beta$ and 
$\gamma$ have been extensively studied in the case $G=SU(3)$, see e.g. 
Ref.~\cite{Nakagawa:2011ar,Klassen:1998ua}. For $SU(2)$ we are only
aware of one study from the literature \cite{Ishiguro:2001jd}. In both cases 
non-perturbative effects turn out to be quite large, so that the analytic
one-loop calculations for $\eta$, which are in principle 
available  for any $N_c$, $\beta$ and $\xi$ \cite{Burgio:2003in}, cannot be 
trusted for practical applications \cite{Klassen:1998ua,Burgio:2003nk}.
 
Since we will concentrate on the case of two colours, $N_c=2$, 
we have decided 
to re-determine the relevant parameters independently by imposing rotational 
invariance for the static potential extracted from space- and time-like Wilson 
loops \cite{Klassen:1998ua,Burgio:2003nk}. Our best estimates for $\gamma$ and 
$a_s$ are given in Tables~\ref{tab:gamma}~and~\ref{tab:as} for a selection of 
values for $\beta$ and $\xi$. These results have been obtained on 
$L^3 \times (\xi\, L)$ lattices of 
spacial extension up to $L=32$ and anisotropies up to $\xi=4$; we have sampled 
$\mathscr{O}(1000)$ configurations using 
a heath-bath algorithm combined with over-relaxation, which for $SU(2)$ can be 
easily extended to the anisotropic case. The same algorithm was also used 
to generate the configurations on which the measurement of Green's functions 
described in the next section have been performed. 

The results for $\gamma$ are mostly compatible within errors with those of 
Ref.~\cite{Ishiguro:2001jd}, while we find some discrepancies in
the scale $a_s$. We have checked that our predictions for $a_s = a_t$ in 
the isotropic case $\xi=1$ agree with others in the literature 
(cfr.~e.g.~Ref.~\cite{Bloch:2003sk}), while the results of 
Ref.~\cite{Ishiguro:2001jd} are always higher than ours at the lower end 
of the scaling window $\beta \lesssim 2.3$. Such discrepancies could be due to 
different systematics inherent to the method used; higher precision might be 
needed to settle the issue.
\begin{table*}[htb]
\begin{center}
\mbox{
\begin{tabular}{|r||c|c|c|c|c|c|c|}
\hline
$\beta$ & 2.15 & 2.2 & 2.3 & 2.4 & 2.5 & 2.6 & 2.7 \\
\hline\hline
$\xi=2$ & 1.654(3) & 1.672(3) & 1.712(3) & 1.754(4) & 1.796(5) & 1.835(6) & 1.870(9)\\
\hline
$\xi=3$ & 2.375(3) & 2.407(3) & 2.474(4) & 2.545(4) & 2.608(5) & 2.663(7) & 2.710(9)\\
\hline
$\xi=4$ & 3.106(4) & 3.151(4) & 3.243(5) & 3.331(5) & 3.406(6) & 3.466(7) & 3.511(9)\\
\hline
\end{tabular}
}
\end{center}
\caption{Bare anisotropy $\gamma$ vs. true anisotropy $\xi$,
for various couplings $\beta$ in $SU(2)$.}
\label{tab:gamma}
\end{table*}
\begin{table*}[htb]
\begin{center}
\mbox{
\begin{tabular}{|r||c|c|c|c|c|c|c|}
\hline
$\beta$ & 2.15 & 2.2 & 2.3 & 2.4 & 2.5 & 2.6 & 2.7 \\
\hline\hline
$\xi=1$ & 1.196(6) & 1.061(6) & 0.821(6) & 0.616(6) & 0.441(5) & 0.290(5) & 0.231(5)\\
\hline
$\xi=2$ & 1.355(7) & 1.206(6) & 0.940(6) & 0.711(6) & 0.511(5) & 0.338(5) & 0.268(5)\\
\hline
$\xi=3$ & 1.391(7) & 1.239(7) & 0.968(6) & 0.732(6) & 0.526(5) & 0.345(5) & 0.270(5)\\
\hline
$\xi=4$ & 1.406(8) & 1.254(7) & 0.979(6) & 0.739(6) & 0.530(5) & 0.346(5) & 0.271(5)\\
\hline
\end{tabular}
}
\end{center}
\caption{Spatial lattice spacing $a_s$ for the choices of $\gamma$ as
in Table~\ref{tab:gamma}. Data are in GeV$^{-1}$, assuming a Wilson string 
tension of $\sigma=$ (440 MeV)$^{2}$. The first line gives our best estimate for
the isotropic scale, cfr.~e.g.~Ref.~\cite{Bloch:2003sk}.}
\label{tab:as}
\end{table*}

It should be noticed that the scale $a_s$ raises with $\xi$ at a fixed coupling 
$\beta$, up to $\xi \gtrsim 3$ where a plateau is reached. In order to simulate 
at the 
same physical point for observables involving spatial links only (like the 
static correlators 
we are interested in), one thus needs to tune $\beta$ to $\xi$ as the 
latter increases.

\subsection{Coulomb gauge and the lattice Hamiltonian limit}
\label{sec:fun}

In Ref.~\cite{Burgio:2008jr,Burgio:2008yg}, it was shown how to quantify and 
treat 
the explicit $a_t$ dependence appearing in static observables after integrating 
their non-static counterparts w.r.t~the energy $p_0$. However, there are still 
more subtle effects of the finite time resolution in lattice simulations. In 
Ref.~\cite{Morningstar:1999rf} it was shown that the lattice Yang-Mills 
spectrum 
suffers from discretization effects of order $\mathscr{O}(a_t^2)$ up to 
$\mathscr{O}(a_t^4)$, depending on the quantum numbers at hand. Such effects 
are 
purely dynamical and linked to the rate at which the theory on an anisotropic 
lattice
reaches the physical Hamiltonian limit $\xi\to\infty$ in which the eigenstates 
and 
the spectrum are well defined. Lattices with $\xi=5$ were eventually used in 
Ref.~\cite{Morningstar:1999rf} to extract the masses of the physical states.

The goal of our investigation is to compare Coulomb gauge lattice results with 
continuum Hamiltonian calculations. Discretization effects at least of 
the same order as for the glueballs, if not larger, should therefore not come 
as a surprise. Let us illustrate this effect using the Coulomb gauge functional 
as an example.

To fix Coulomb gauge on each MC generated configuration we adapt the algorithm 
of Ref.~\cite{Bogolubsky:2005wf,Bogolubsky:2007bw} as described in 
Ref.~\cite{Burgio:2008jr,Burgio:2008yg,Burgio:2009xp}, i.e. we first fix one of 
the $N_c^d = 2^3 = 8$ center flip sectors and then maximize, via a simulated 
annealing 
plus ensuing over-relaxation, separately at each time slice $t$ the gauge 
functional
\beq 
F(t) = \frac{1}{3 N_c L^3} \sum_{\vek{x},i} \mbox{Tr} ~
U^g_{i}(\vek{x},t)\;,\qquad
U^g_{i}(\vek{x},t) = g(\vek{x},t) ~U_{i}(\vek{x},t) 
~g^{\dagger}(\vek{x}+{\hat{i}},t)
\label{eq:functional}
\eeq
with respect to local gauge transformations $g(\vek{x},t) \in SU(2)$.
To further improve the gauge fixing quality, we start the gauge fixing engine 
with $n_c$ random 
gauge copies of the initial time slice (with usually $5 \le n_c \le 10$), and 
select
the gauge-fixed copy with the highest value of $F(t)$. We then combine these 
best gauge-fixed time slices into a best gauge-fixed configuration for the 
flip sector chosen. Finally, we proceed to the next sector and repeat the 
procedure, 
selecting in the end the sector which gave the highest global functional 
$F = \sum_t F(t)$. Every time slice in the final gauge-fixed copy 
for each configuration has therefore been selected among at least 
$n_c \cdot N_c^d = 40\ldots80$ 
gauge fixing runs with random starting points from all topological sectors. 
On top of this elaborated procedure to fix the spatial gauge freedom in each 
time slice separately we fix the residual temporal gauge symmetry
$g(t)$ via the integrated Polyakov gauge, as described 
in Ref.~\cite{Burgio:2008jr,Burgio:2008yg,Burgio:2009xp}.

\begin{figure}[htb]
\subfloat[][]{\includegraphics[width=0.45\textwidth,height=0.4\textwidth]{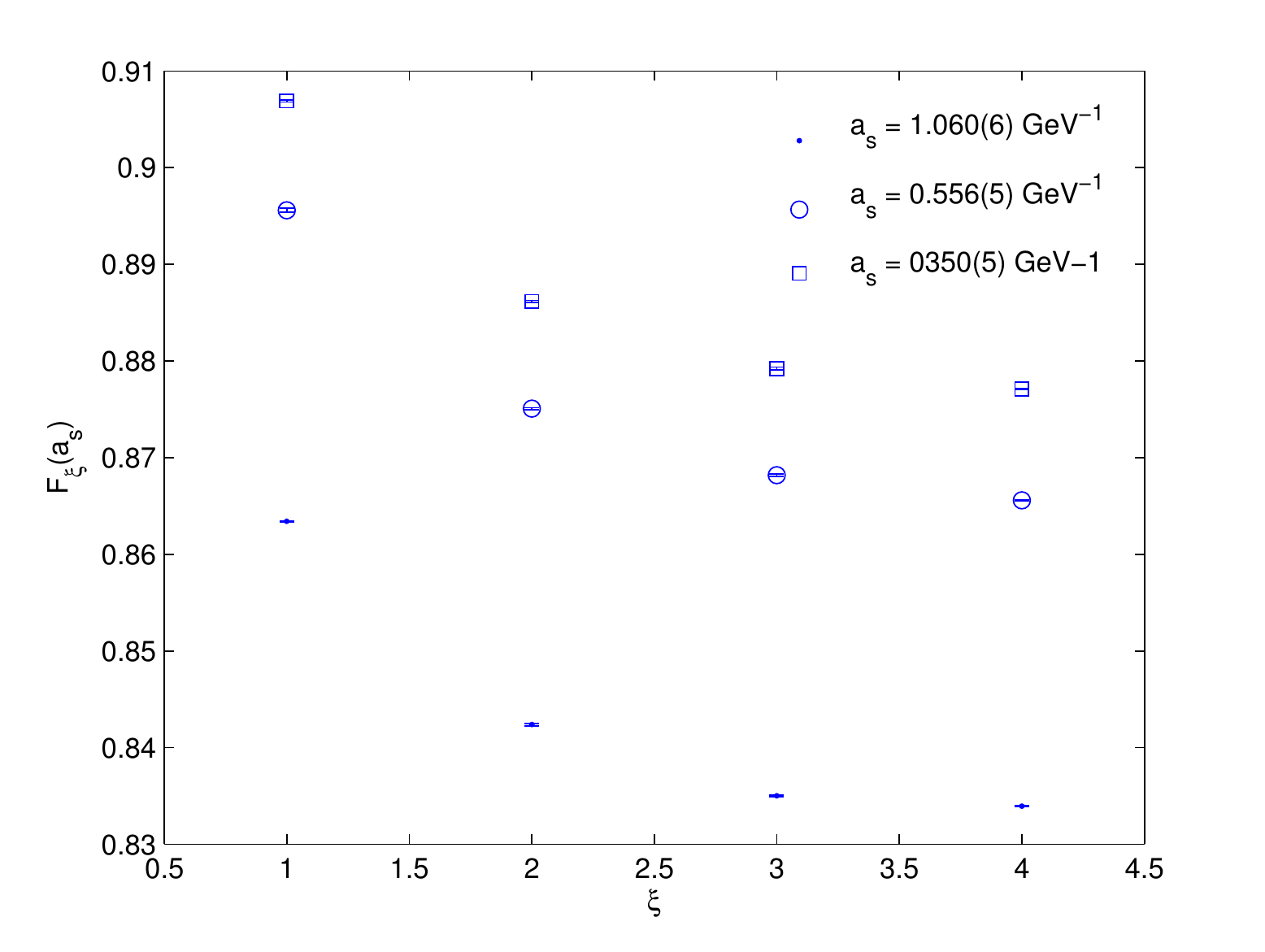}}
\subfloat[][]{\includegraphics[width=0.45\textwidth,height=0.4\textwidth]{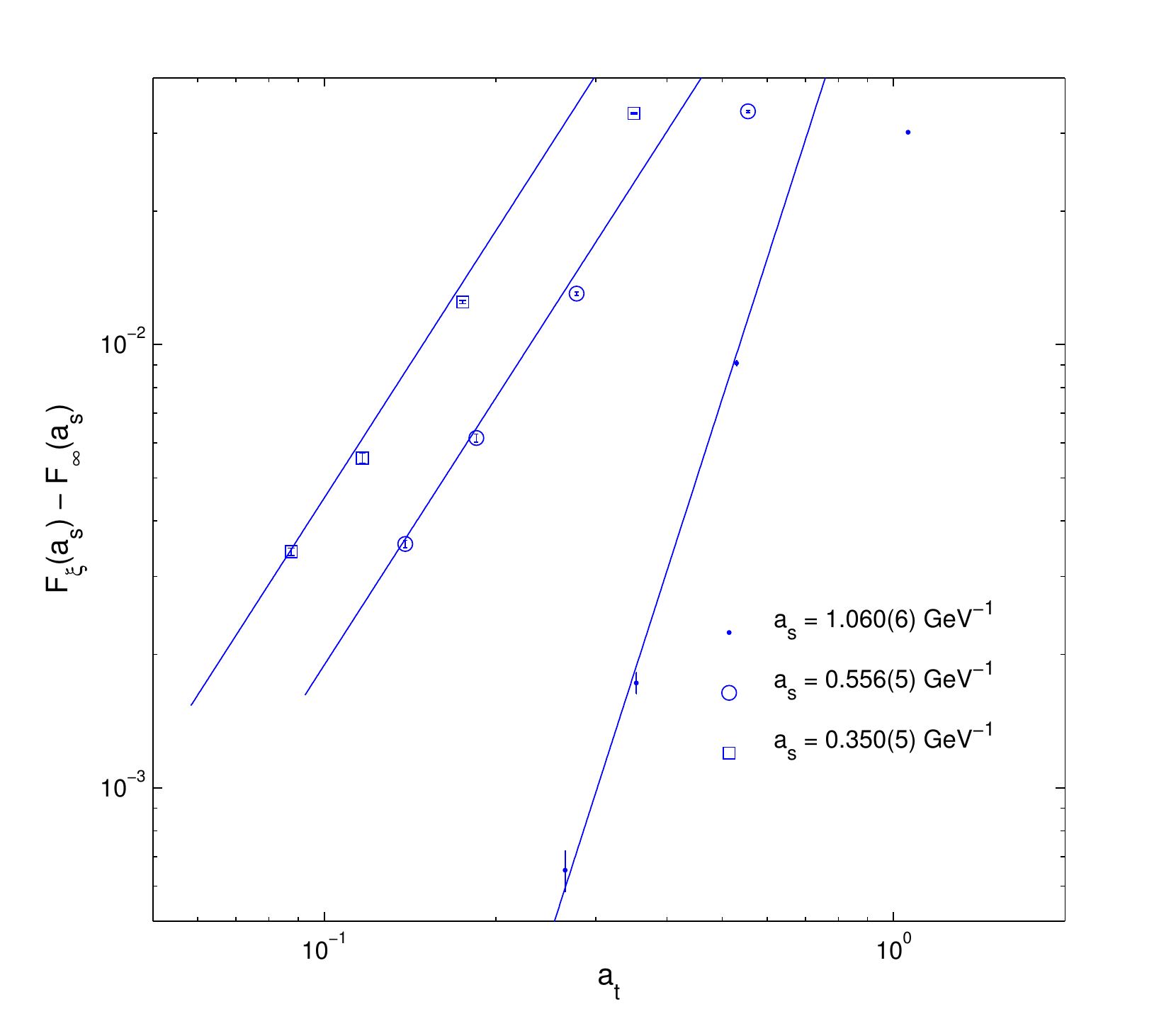}}
\caption{{(a)}: dependence of the gauge fixing functional $F_\xi(a_s)$ 
on the anisotropy $\xi$ at fixed spatial cut-off $a_s$. {(b)}: Deviation 
of the gauge fixing functional from the Hamiltonian limit,
$F_\xi(a_s) - F_{\infty}(a_s)$, as a function of the temporal lattice spacing $a_t$,
 together with its leading power corrections.}
\label{fig1}
\end{figure}

As we have seen in the previous section, a constant spatial cut-off $a_s$ 
for increasing anisotropy $\xi$ can be obtained by tuning $\beta$ to $\xi$; 
$a_t = \xi^{-1}\,a_s$ will then decrease linearly with $\xi^{-1}$.
Naively one would then expect all observable which only 
depend on the spatial links to be independent of $\xi$. For instance, 
the Coulomb gauge functional Eq.~(\ref{eq:functional}) should, in principle, 
be a function of $a_s$ only, at least if our algorithm is good enough to 
approach the absolute minimum reliably, and if finite volume effects can 
be ignored \footnote{Notice that the complexity of our gauge fixing 
algorithm scales with the spatial volume $L^3$ since all time slices are 
fixed independently.}. This is not the case, as can be seen from 
Figure~\ref{fig1}: the left chart displays the best value of the gauge 
functional $F$ from our algorithm as a function of the anisotropy $\xi$, 
for three fixed values of the lattice spacing 
$a_s =$ 1.060(6), 0.556(5) and 0.350(5) GeV$^{-1}$. 
As the Hamiltonian limit is approached by increasing $\xi$, the functional 
$F$ decreases, i.e.~the configurations at fixed time slice become 
``rougher'' 
although $a_s$ is kept constant; as can be seen from the figure, 
the corrections with $\xi$ are several orders of magnitude higher than the 
Gribov noise. The leading
order in the corrections to $F_\xi$ are well described by a power law in 
$\xi^{-1} \propto a_t$. Our best estimates for the asymptotic values $F_{\infty}$
together with the coefficient of the leading order corrections are given in 
Tab.~\ref{tab:F}, while the right chart of Fig.~\ref{fig1} shows 
$F_\xi - F_{\infty}$ 
versus the leading correction in $a_t$. For the stronger couplings, this 
leading correction is $\mathscr{O}(a_t^4)$, while for weaker couplings we find 
it 
to scale as $\mathscr{O}(a_t^2)$.
\begin{table*}[htb]
\begin{center}
\mbox{
\begin{tabular}{|r||c|c|c|}
\hline
       & $F_{\infty}$ & $c_2$ & $c_4$ \\
\hline\hline
$a_s=1.060(6)$ GeV$^{-1}$ & 0.8333(2) & $--$ & 0.152(1)\\
\hline
$a_s=0.556(5)$ GeV$^{-1}$ & 0.8620(2) & 0.058(1) & $--$\\
\hline
$a_s=0.350(5)$ GeV$^{-1}$ & 0.8737(2) & 0.055(1) & $--$\\
\hline
\end{tabular}
}
\end{center}
\caption{Asymptotic extrapolation of $F_\xi$ together with the
coefficient of the leading power correction of oder $\mathscr{O}(a_t^4)$ 
($a_s=1.060(6)$ GeV$^{-1}$) and $\mathscr{O}(a_t^2)$ 
($a_s=0.556(5)$ and $0.350(5)$ GeV$^{-1}$).}
\label{tab:F}
\end{table*}

This sensitivity to $\xi$ (or $a_t$) does not only occur in the gauge fixing
functional. In Ref.~\cite{Burgio:2008yg,Nakagawa:2009is,Nakagawa:2011ar} a 
strong 
dependence on the anisotropy was also seen for the temporal gluon propagator 
Eq.~(\ref{tprop}), although this was numerically shown to be energy 
(i.e. $U_0$) 
independent \cite{Burgio:2008jr}. Also,
preliminary results indicate that the Gribov mass $M$ in Eq.~(\ref{gribov})
slightly increases with $\xi$, saturating to a value $M\simeq 1.2(2)$
for $\xi = 3 - 4$.
In Sec.~\ref{sec:ghost} and 
\ref{sec:potential} we 
will see that the ghost propagator Eq.~(\ref{stgh}) and the Coulomb 
potential 
Eq.~(\ref{stpot}), which are by definition independent of the time-like 
links, both turn out to be sensitive to the anisotropy.

From Fig.~\ref{fig1} one can infer that lattices with $\xi \gtrsim 6$ would be 
needed to minimize the $a_t$ corrections in $F$, i.e. to reach the plateau
at $F_{\infty}$. Since we also need a large spatial extension $L\simeq 50$ to 
explore the deep infrared momentum region, the combination of these two 
requirements would force us to simulate on lattices of temporal extension 
$N_t \simeq 300$, beyond the computational power at 
our disposal. Moreover, not all observables
need to be as sensitive to $\xi$ as the gauge functional. We will therefore 
restrict our investigation to $\xi=1\ldots 4$ and attempt to extrapolate to 
larger $\xi$ wherever necessary. Of course, a direct confirmation of our 
results on large lattices with high anisotropies would be highly desirable.

\subsection{Ghost Form Factor}
\label{sec:ghost}
The ghost form factor in Coulomb gauge, Eq.~(\ref{stgh}), has been discussed in 
Refs.~\cite{Langfeld:2004qs,Nakagawa:2007fa,Nakagawa:2009zf,Burgio:2010wv}; 
neither its ultraviolet nor its infrared behavior could be determined 
conclusively.
In the UV, the primary goal is to check the sum rule for the  anomalous 
dimensions in 
Eq.~(\ref{sumrules}). In the IR, the main question is whether the ghost
propagator is compatible with an infrared finite behaviour, as is the case in 
Landau gauge \cite{Bogolubsky:2009dc}. If this were true, it would of course 
spoil 
the Gribov-Zwanziger mechanism and the dual superconductor argument of 
Ref.~\cite{Reinhardt:2008ek}.

We calculate $d(\vek{p})$ by inverting the Coulomb gauge Faddeev-Popov (FP) 
operator through a conjugate gradient algorithm on lattices of spatial sizes 
up to 
$54^3$ and anisotropies up to $\xi=4$. Although for most configurations the 
algorithm 
works quite well, there are some "exceptional" cases where very small 
eigenvalues of the 
FP operator make the inversion ill-conditioned, signaled by very bad 
convergence. 
On the other hand, we expect exactly these configurations to contribute 
substantially 
to the infrared divergence of the ghost form factor, since they lie close to 
the 
Gribov horizon. Our current procedure is to exclude such configurations from 
the Monte-Carlo ensemble; this obviously creates a bias in the data which 
potentially suppresses the ghost propagator at very low momenta. 

Whether or not these 
near-horizon configurations have a measurable effect on the ghost propagator 
also depends on the frequency with which they appear within a MC sequence;
contrary to Landau gauge \cite{Bogolubsky:2005wf}, 
this seems to be relatively stable upon improvement of the gauge
fixing. In our studies we have observed an ill-defined FP operator in about 
one of 
every 
$300$ configurations, with very large uncertainties (i.e.~there were also MC 
runs 
with over $1000$ configurations that did not show a single near-horizon 
configuration).
Preliminary studies of individual singular configurations with improved 
pre-conditioners
and numerics indicate that their contribution to the ghost form factor can be 
enhanced by up to a factor $40$ as compared to the ensemble average, again 
with large 
uncertainties due to the bad condition number. Clearly, this is a zero measure 
times
infinite contribution problem that can only be decided by much improved 
statistics 
combined with better inversion algorithms. In this study we have not been able 
to resolve
this issue quantitatively; the exponent which we will extract in the 
following 
must therefore be considered as a \emph{lower bound} for the correct infrared 
behaviour. 

\begin{figure}[htb]
\subfloat[][]{\includegraphics[width=0.45\textwidth,height=0.4\textwidth]{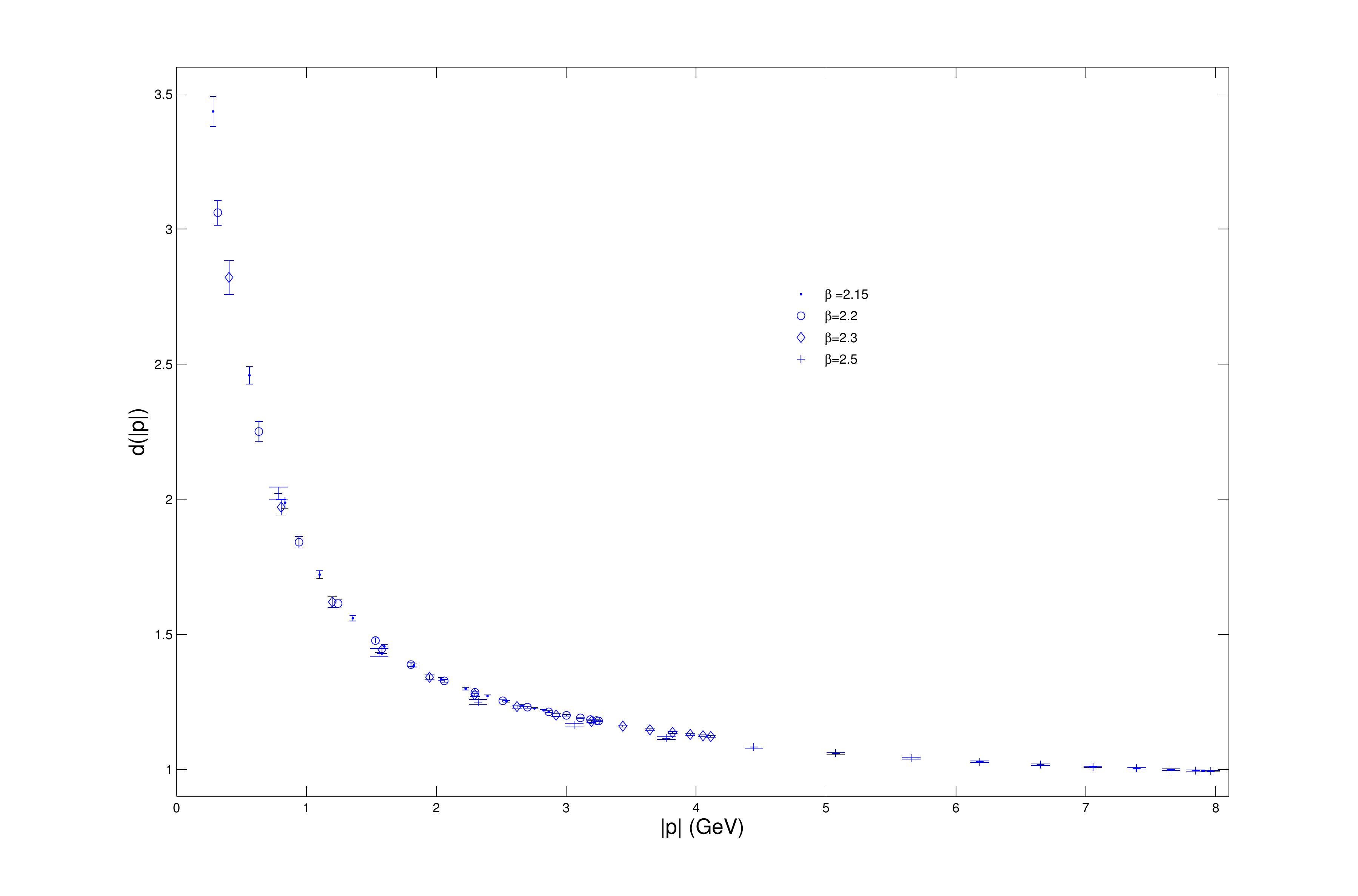}}
\subfloat[][]{\includegraphics[width=0.45\textwidth,height=0.4\textwidth]{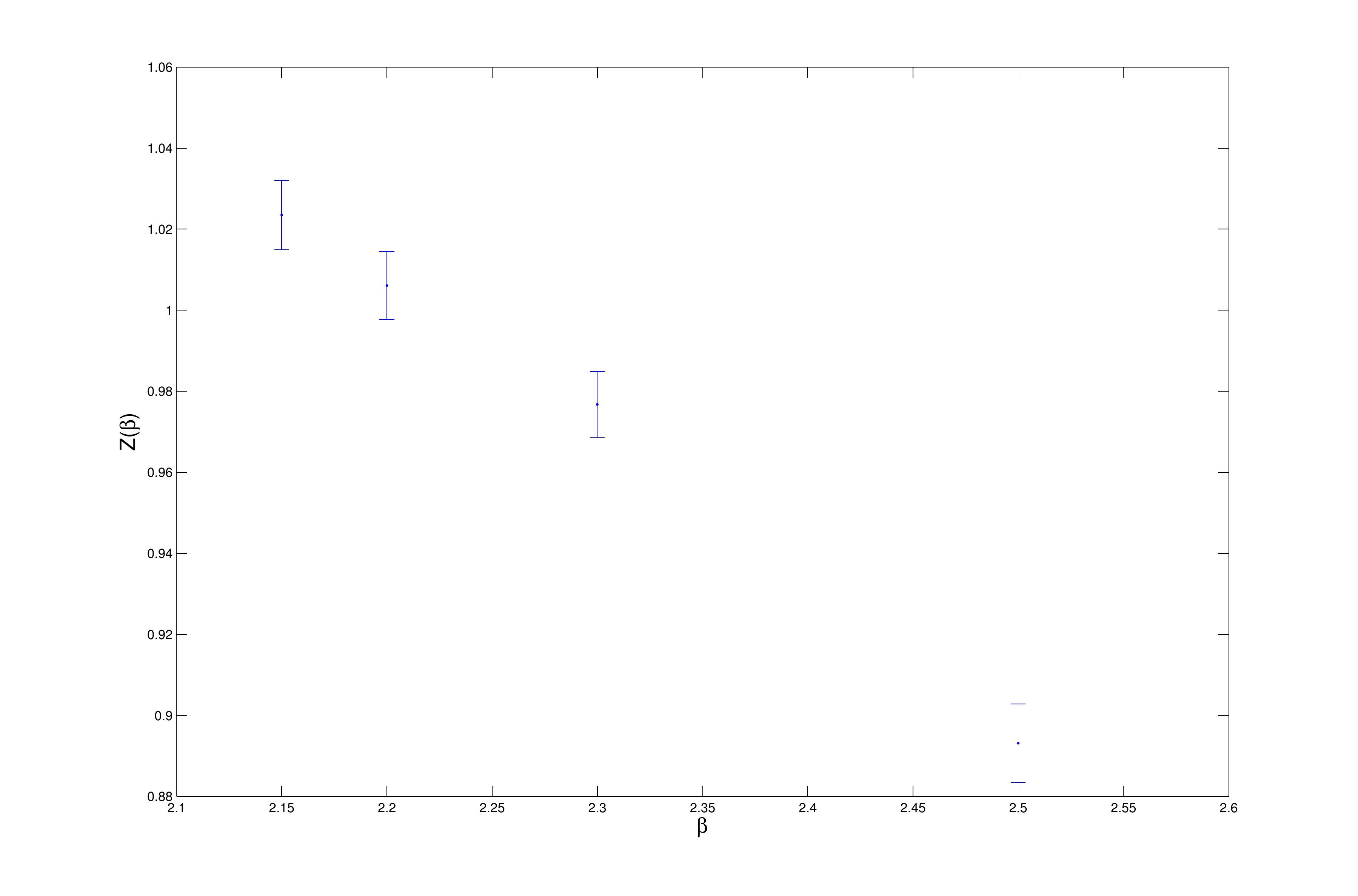}}
\caption{{(a):} Ghost form factor $d(|\vek{p}|)$ for anisotropy 
$\xi=1$. {(b):} Renormalization coefficients for the ghost propagator, 
as a function of the coupling $\beta$.}
\label{fig2}
\end{figure}

\begin{figure}[htb]
\subfloat[][]{\includegraphics[width=0.45\textwidth,height=0.4\textwidth]{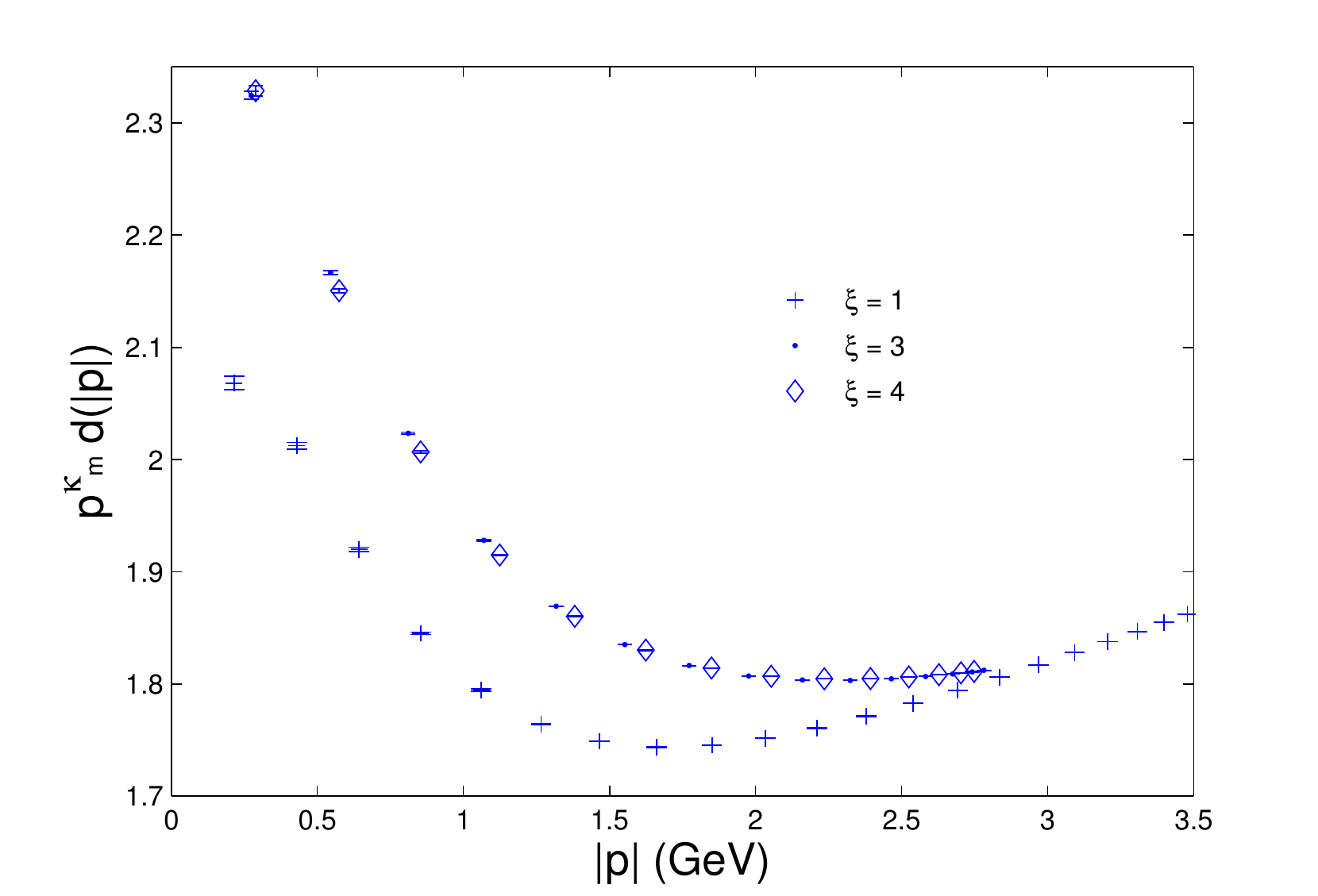}}
\subfloat[][]{\includegraphics[width=0.45\textwidth,height=0.4\textwidth]{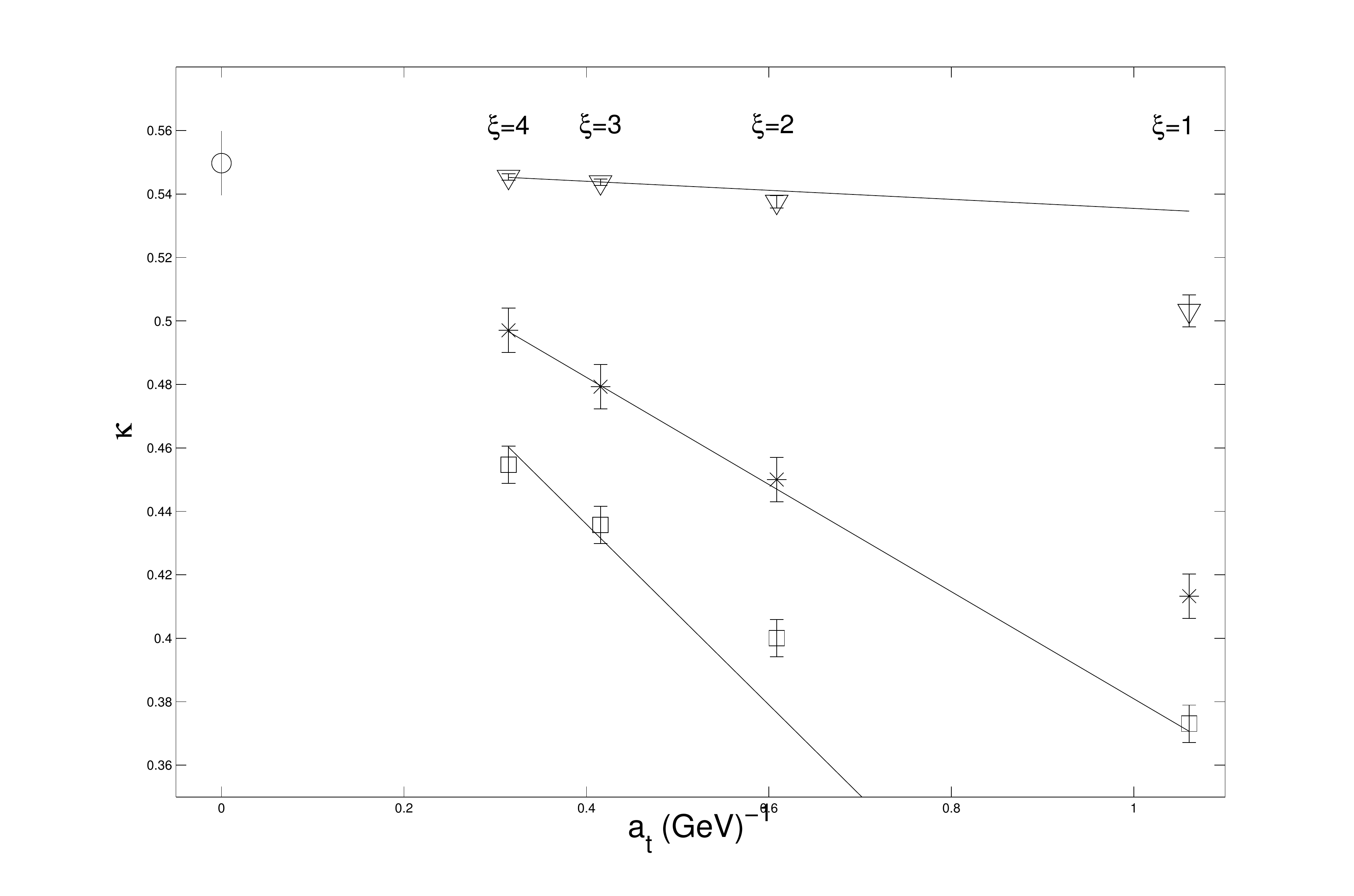}}
\caption{{(a)} Infrared behavior of $|\vek{p}|^{\kappa_m} \, d(\vek{p})$ 
for different anisotropies $\xi$. {(b)} $\kappa$ extrapolation for
$a_t\to 0$; Different symbols correspond to different fitting methods to
estimates $\kappa$; points at fixed $a_t$ ($\xi$) correspond to the same 
data set for $d(|\vek{p}|)$; lines give the leading correction 
$\propto a_t$ to $\kappa_{\rm gh}$.}
\label{fig3}
\end{figure}

Results for  $d(\vek{p})$ at $\xi=1$ are shown in Fig.~\ref{fig2}(a), while in 
Fig.~\ref{fig2}(b) we give the coefficients $Z(\beta)$ needed to 
multiplicatively
renormalize the data at different coupling, arbitrarily scaled such to let 
$d(\vek{p}) \simeq 1$ at large momentum.
Although at first glance the data agree qualitatively with the above 
expectations, on a closer look two problems appear. First, small 
scaling violations can be measured within errors in the ultra-violet region; 
a fit to a logarithmic behavior is thus afflicted with large errors. 
Second, the infrared region shows two different behaviors for an 
intermediate momentum range
$0.5$~GeV$ \lesssim |\vek{p}| \lesssim 1.2$ GeV 
and a low momentum range 0.2 GeV $\lesssim |\vek{p}| \lesssim 0.5$ GeV. In the 
first regime the data are compatible with a power-law behavior 
\begin{equation}
d(|\vek{p}|) \propto |\vek{p}|^{-\kappa}, 
\end{equation}
with $\kappa \simeq 0.5$, while in the low-momentum region the effective
$\kappa$ decreases, approaching a value $\kappa\lesssim 0.4$. By going to 
higher anisotropies 
this behavior softens and the data become more and more compatible with a
stable power-law. To illustrate this effect, in Fig.~\ref{fig3}(a) we 
highlight the infrared behavior showing 
$|\vek{p}|^{\kappa_m} d(\vek{p})$ for different anisotropies, where 
$\kappa_m = 0.373(6)$ is the effective exponent we have measured for 
$L=54$, $\xi=1$ in the lowest momentum region. 
The data for $\xi=1$ go to a 
constant while for higher anisotropies a power-law still describes the data 
well. 
In Fig.~\ref{fig3}(b) we show the IR exponents $\kappa$ obtained from three
different fitting methods for each 
set of data at fixed anisotropy $\xi$, i.e.~temporal cut-off $a_t$: 
the lower curve corresponds to a pure 
$|\vek{p}|^{-\kappa}$ behaviour cutting the data at $|\vek{p}|\lesssim 1.5$ GeV;
the middle one adds a linear subleading correction and leaves the same cut;
the higher one combines the linear correction to the power law with a cut
at $|\vek{p}|\lesssim 1$ GeV.
Fitting all these values of $\kappa$ to a constant $\kappa_{\rm gh}$, constrained
to be the same for all three methods used, plus power law corrections in $a_t$ 
we find
\begin{equation}
\kappa_{\rm gh} = 0.55(1) 
\label{eq:ir_ghost}
\end{equation}
with leading correction of order $\mathscr{O}(a_t)$ and a $\chi^2$/d.o.f. 
$\simeq 0.3$.

Because of the cut on
exceptional configurations discussed above we have of
course introduced a bias in our data. Therefore the value in 
Eq.~(\ref{eq:ir_ghost}) can only be considered as a lower bound on the actual
value of $\kappa_{\rm gh}$. Still, as far as we know, this is the first time
that an IR enhanced ghost form factor can be reliably proven to exist on the 
lattice, making the Gribov-Zwanziger confinement scenario
{\it in Coulomb gauge} 
fully consistent. Indeed, while in Ref.~\cite{Burgio:2009xp} it was shown that 
the Gribov gluon propagator Eq.~(\ref{gribov}) is IR equivalent to the 
massive Landau gauge solution, in the latter case lattice simulations
always give an IR finite, i.e. tree-level like ghost form factor. How this
can be made to agree with the Gribov-Zwanziger mechanism is still a debated
issue.

In general, although \emph{any} divergence ($\kappa_{\rm gh} > 0$) of the ghost 
form factor would be sufficient to support
the Gribov-Zwanziger mechanism, the exact value of the 
infrared exponent $\kappa_{\rm gh}$ matters for the sum rule 
Eq.~(\ref{sumrules}), which is violated by our findings combined with the 
results for Gluon propagator in 
Refs.~\cite{Burgio:2008jr,Burgio:2008yg,Burgio:2009xp}. There are 
several possible resolutions to this issue: first, our data for the 
ghost propagator only covers the range down to 
$|\vek{p}| \simeq 200\,\mathrm{MeV}$
with sufficient statistics and precision. Further changes in $\kappa_{\rm gh}$
at a much smaller scale $\lambda_{IR} \ll \Lambda_{QCD}$ can not be ruled out, 
although we currently do not see 
the onset of such deviations, while the appearance of an additional small 
scale in pure YM theory would raise a different kind of interpretation problem.
Moreover, as we have explained above, our data for $\kappa_{\rm gh}$ is only 
a lower bound since we have neglected singular configurations from near the 
Gribov horizon. Whether or not these configurations contribute substantially to 
the ghost propagator and its exponent $\kappa_{\rm gh}$ cannot be predicted 
with our current computational resources. Most likely, it would also require 
algorithmic 
changes in the inversion method for ill-conditioned FP operators, and maybe 
even 
a change in the fundamental MC generator to better sample the near-horizon 
region in field space.   
 
Besides these caveats at our numerical data, the origin of the sum rule
Eq.~(\ref{sumrules}) itself leaves room for discussion. In essence, it is based 
on the assumption that (i) the gluon and ghost propagators have conformal
(power-like) behaviour in the infrared and (ii) the ghost-gluon vertex receives 
no radiative corrections (other than an overall multiplicative renormalization) 
at low momenta, i.e.~it is essentially \emph{bare}. 
Both results are borrowed from Landau gauge, where Taylor's theorem gives a 
firm explanation why the ghost-gluon vertex is trivial as $p \to 0$. The 
ensuing sum rule for gluon and ghost propagator are then simple consequences, 
at least as long as massive solutions in the infrared can be ruled out, for 
which the ghost and gluon behaviour would \emph{decouple}. All these 
assumptions 
are well confirmed by lattice simulations in Landau gauge. 

For Coulomb gauge, however, the situation may be more involved. A careful 
analysis of the 
Slavnov-Taylor identities in Coulomb gauge exhibit a complicated interplay
between transversal (spatial) and longitudinal (temporal) degrees of freedom,
see Eq. (4.12) in Ref.~\cite{Watson:2008fb}. 
Integrating over all energies $p_0$ to reach the equal-time limit could 
therefore
induce non-trivial structure in the \emph{static} Green functions, even 
though the proof of Taylor's theorem carries over to Coulomb gauge
in the limit $|\vek{p}| \to 0$ at any fixed $p_0$. Unfortunately, no 
analytical calculations of the \emph{static} vertex at low momenta based on the 
Slavnov-Taylor identities has so far been possible, 
and the corresponding calculation on the lattice have not been carried out.
In Ref.~\cite{Campagnari:2011bk}, within the Hamiltonian approach, the 
Dyson-Schwinger equation for the ghost-gluon vertex was solved at the one
loop level and little dressing was found. Furthermore, 
Landau gauge calculations in three (euclidian) dimensions also exhibit little 
dressing of the ghost-gluon vertex; it is however unclear if these results
simply carry over to the four-dimensional {\em static} quantities in Coulomb 
gauge.

\begin{figure}[htb]
\includegraphics[width=0.8\textwidth,height=0.65\textwidth]{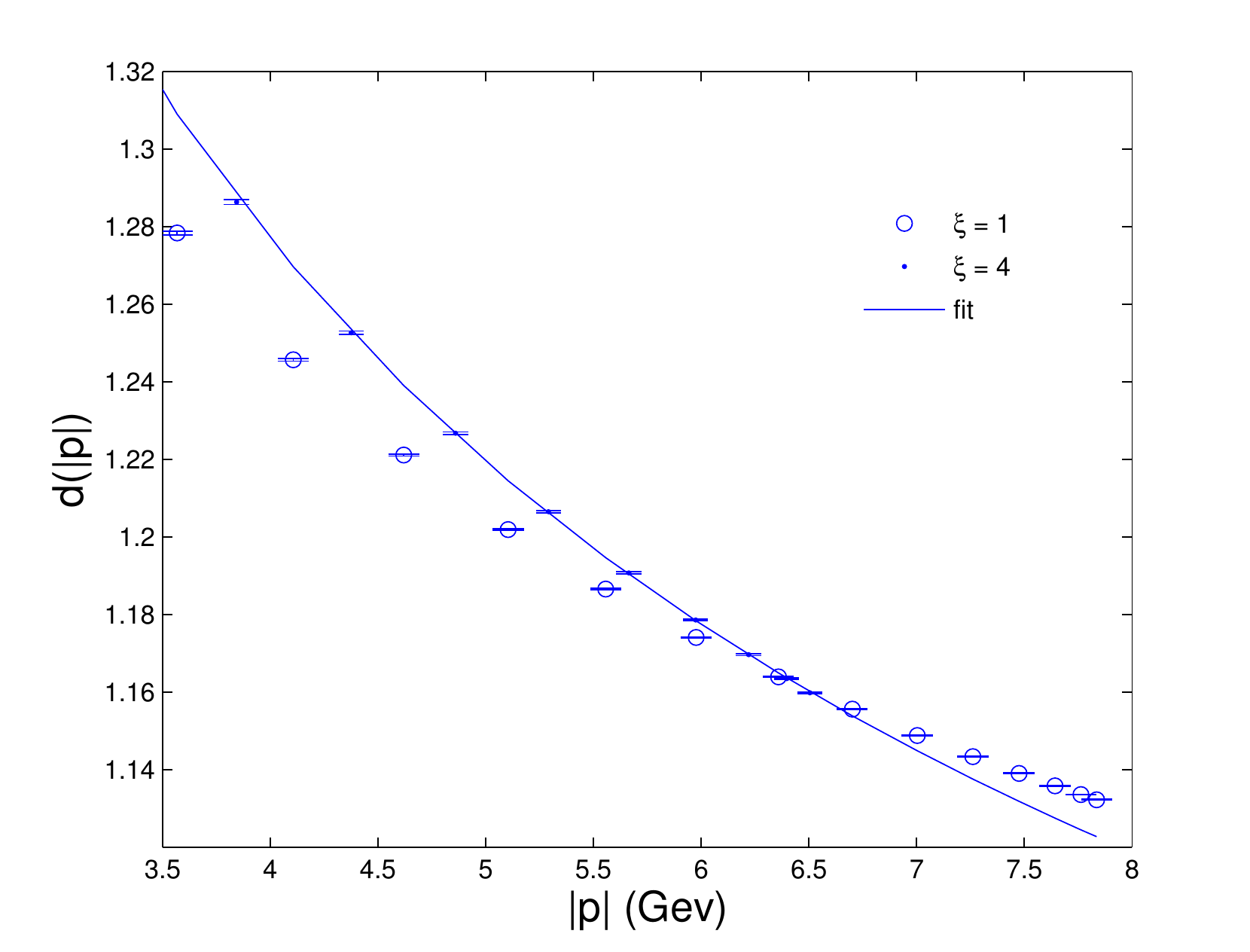}
\caption{UV behavior of $d^{-1}(\vek{p})$ for different 
anisotropies $\xi$ compared with Eq.~(\ref{UVexp}).}
\label{fig4}
\end{figure}

Let us now consider the ultra-violet behavior, where exceptional configurations 
play no role and the situation is much clearer. 
As mentioned above, slight scaling violations can be measured in the 
data for $\xi=1$. These diminish as $\xi$ is increased and the anomalous 
dimension 
of the ghost field can be determined in the Hamiltonian limit. 
In Fig.~\ref{fig4} we compare $d(\vek{p})$ for the lowest and highest
anisotropies $\xi$.  For the latter we show our best fit to the expected
asymptotic behaviour with $\gamma$ given by the sum-rule 
in Eq.~(\ref{NPexp}); the ultra-violet mass scale from 
the fit is $m = 0.21(1)$~GeV, cfr.~Eq.~(\ref{UVexp}).

\subsection{Coulomb Potential}
\label{sec:potential}
The Coulomb potential, Eq.~(\ref{stpot}), has been intensively investigated in the 
literature 
\cite{Greensite:2003xf,Nakamura:2005ux,Nakagawa:2006at,Voigt:2008rr,%
Nakagawa:2008zza,Nakagawa:2009is,Nakagawa:2010eh}.
While there is general agreement that its infrared behavior is determined by 
a Coulomb string tension larger than the Wilson string tension 
\cite{Zwanziger:2002sh}, the value quoted for the ratio $\sigma_C/\sigma$ 
varies among the above works. 

\begin{figure}[htb]
\includegraphics[width=0.8\textwidth,height=0.65\textwidth]{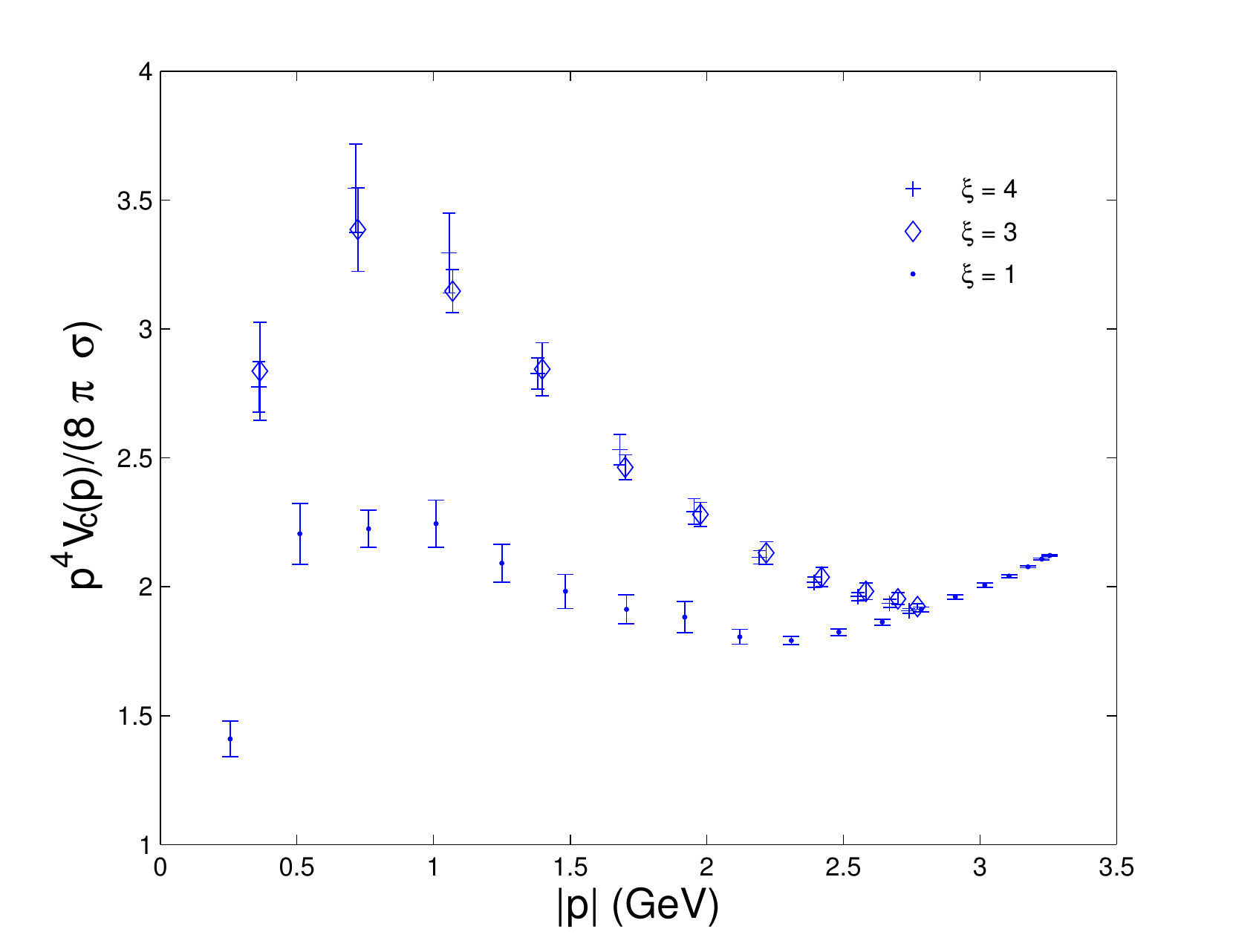}
\caption{Infrared behavior of $|{\vek{p}}|^4\, V_C(\vek{p})/(8 \pi \sigma)$ for 
different anisotropies $\xi$.}
\label{fig5}
\end{figure}

We have calculated the Coulomb potential Eq.~(\ref{stpot}) for lattices 
of spatial extension up to $40^3$ and anisotropies up to $\xi=4$. Exceptional
configurations did not play a major role in this case; this may be 
either due to the smaller volumes and lower statistics employed, since the
computation of Eq.~(\ref{stpot}) is much more expensive than Eq.~(\ref{stgh}),
or to the better pre-conditioning due to the Laplace inversion, or both. 
In Fig.~\ref{fig5} we show the infrared behavior of 
$|{\vek{p}}|^4\, V_C(\vek{p})/(8 \pi \sigma)$ for different values of $\xi$;
the reason for the normalization will be clear in the following. 
As in the case of the ghost form factor, we observe a change in the data 
in the Hamiltonian limit. The corrections due to the 
finite temporal resolution seem to saturate at our highest values of $\xi$,
although error bars are still quite large. 
Also, the bending down of the data at $\simeq$ 0.5 -- 1 GeV, observed
in all above quoted analyses for $\xi=1$, becomes 
milder, although it does not disappear. Its presence should indeed not come as
a surprise: at least for the physical potential $V(r)$, which we know to be 
given by
\cite{Luscher:1980fr}
\begin{equation}
V(r) = \sigma r + \mu -\frac{\pi}{12 r} + \mathscr{O}(\frac{1}{r^2})\,,
\label{l-trm}
\end{equation} 
we expect, after introducing a suitable IR cutoff $\lambda$ and 
taking $\lambda \to 0$ \footnote{A Yukawa
fall-off $e^{- \lambda r}$ is convenient to perform 
the Fourier integrals, but any other IR cutoff, e.g. a finite lattice 
size, will work as well.
Of course, the two limits $\lambda \to 0$ and $|{\vek{p}}|\to 0$ do not 
commute; the thermodynamic limit $\lambda \to 0$ must be taken first}:  
\begin{equation}
|{\vek{p}}|^4\, V(\vek{p})  = 8 \pi \sigma + \frac{\pi^2}{3} |{\vek{p}}|^2
+ \mathscr{O}(|{\vek{p}}|^{3})\,,
\label{l-term}
\end{equation}
since the contribution from $\mu$ in Eq.~(\ref{l-trm}), which gives 
a term $\propto |{\vek{p}}|$ in Eq.~(\ref{l-term}), will vanish with 
$\lambda$. Thus $|{\vek{p}}|^4\, V(\vek{p})$ must have a minimum as 
$|{\vek{p}}|\to 0$; since its asymptotic UV
behaviour is also, up to logarithmic correction, $\propto |{\vek{p}}|^2$, we
only have two possibilities: either $|{\vek{p}}|^4\, V(\vek{p})$ is a monotonic
function of $|{\vek{p}}|$ or, if it has some local minimum for some
$|{\vek{p}}|_m > 0$, then it must have a maximum in 
$0 < |{\vek{p}}|_M < |{\vek{p}}|_m$.

If $V_C$ asymptotically behaves like $\sigma_C\, r$ 
the data in Fig.~\ref{fig5} must approach a 
constant as $|\vek{p}| \to 0$, giving direct access to the Coulomb string 
tension. 
It is natural to parametrize them as in Eq.~(\ref{l-term}):
\begin{equation}
\frac{|{\vek{p}}|^4\, V_C(\vek{p})}{8 \pi \sigma}  = \frac{\sigma_C}{\sigma} 
+ \alpha |{\vek{p}}| + \gamma |{\vek{p}}|^2 + \mathscr{O}(|{\vek{p}}|^{3})\,.
\label{ftc}
\end{equation}
where we would expect $\gamma > 0$ and $\alpha \to 0$ in the 
thermodynamic limit.
Fitting the data to Eq.~(\ref{ftc}) and introducing different cuts 
at different momenta we obtain for the Coulomb string tension 
\begin{equation}
\sigma_C = 2.2(2) \,\sigma\,.
\end{equation}
Notice that, even though we don't have much data in the IR region, if we 
optimistically 
constrain $\alpha \equiv 0$ we obtain a higher value for the Coulomb string 
tension, 
$\sigma_C = 2.5(1) \,\sigma$. 
Better statistics and data from larger lattices and/or anisotropies 
in the low momentum region would of course be welcome to improve the result. 

\section{Conclusions}
We have shown that a sufficient approach to the Hamiltonian limit, 
which on the lattice translates into anisotropic lattices with a high temporal 
resolution, is crucial for a successful investigation of the ghost form factor 
and the Coulomb potential. The effect of the anisotropy on static correlators lies 
in the different dynamics for fixed spatial lattice spacing $a_s$ as $a_t$ decreases, 
as we have shown in Sec.~\ref{sec:fun}. This explains the dependence on $\xi$ 
found also for explicitly energy independent observables that can be evaluated 
in a single time slice.

In Sec.~\ref{sec:ghost} we have shown that the infrared exponent of the 
ghost form factor saturates at $\kappa_{\rm gh} \approx 0.55(1)$ when taking the 
Hamiltonian limit. Although our estimate may only be a lower bound due to 
the contribution of near-horizon configurations, the infrared divergence 
of the ghost form factor confirms the Gribov-Zwanziger \cite{Gribov:1977wm,Zwanziger:1995cv}
confinement mechanism and the vanishing of the dielectric function for the Yang-Mills 
vacuum \cite{Reinhardt:2008ek}. We have also confirmed quantitatively the sum rule 
for the anomalous dimensions of the static gluon and ghost fields in Coulomb gauge.
The similar sum rule prediction in the infrared would require 
$\kappa_{\rm gh} = 1$ in order to be compatible with the Gribov formula confirmed in 
ref.~\cite{Burgio:2009xp}. This is clearly violated by our best fit 
$\kappa_{\rm gh} \approx 0.55(1)$. 
We have discussed possible arguments to resolve this discrepancy, by 
critically analyzing both our data and the origin of the sum rule itself.
A conclusive settlement of this issue must be left to a future investigation. 
Finally, in Sec.~\ref{sec:potential} we have shown 
that the Coulomb potential is linearly rising in position space, and the corresponding 
Coulomb string tension settles nicely in the Hamiltonian limit $\xi \to \infty$, where 
we extract a value of $\sigma_C \simeq 2.2\,\sigma$.

As indicated above, future investigations in this direction should concentrate 
on the ghost form factor and Coulomb potential on much larger spatial lattices 
to probe lower momenta, 
with improved statistics and better numerics to tackle the rare near-horizon 
configurations. This may help to settle the remaining question of the infrared 
sum rule for the exponents in the power-law of the low order Green functions.

\section*{Acknowledgments} 
We would like to thank Peter Watson and Davide R. Campagnari for stimulating 
discussions.  This work was partly supported by DFG under 
the contract DFG-Re856/6-3. 
\bibliographystyle{apsrev}
\bibliography{references}

\end{document}